\documentclass[aps,prl,twocolumn,groupedaddress,showpacs,floatfix,superscriptaddress,longbibliography]{revtex4-1}
\usepackage{epsfig}
\usepackage{amsmath,amssymb}
\usepackage{graphicx}
\usepackage[dvipsnames,usenames]{color}
\usepackage[normalem]{ulem}
\usepackage{hyperref}
\usepackage{soul} 
\emergencystretch=\maxdimen
\begin{document}

\title{Charge-Density Wave Order
on a $\pi$-flux Square Lattice}
\author{Y.-X. Zhang}
\affiliation{Department of Physics, University of California, Davis, CA 95616,USA}
\author{H.-M. Guo}
\affiliation{Department of Physics, Key Laboratory of Micro-nano Measurement-Manipulation and Physics (Ministry of Education),
Beihang University, Beijing, 100191, China}
\author{R. T. Scalettar}
\affiliation{Department of Physics, University of California, Davis, CA 95616,USA}

\begin{abstract}
The effect of electron-phonon coupling (EPC) on Dirac fermions has
recently been explored numerically on a honeycomb lattice, leading to precise
quantitative values for the finite temperature and
quantum critical points.
In this paper, we use the unbiased determinant Quantum Monte Carlo (DQMC)
method to study the Holstein model on a half-filled staggered-flux
square lattice, and compare with the honeycomb lattice geometry,
presenting results for a range of phonon frequencies
$0.1 \leqslant \omega \leqslant 2.0$. We find that the interactions give rise to charge-density
wave (CDW) order, but only above a finite coupling strength
$\lambda_{\rm crit}$. The transition temperature is evaluated and presented
in a $T_c$-$\lambda$ phase diagram. An accompanying mean-field theory (MFT) calculation also predicts the existence of quantum phase transition (QPT), but at a substantially smaller coupling strength.
\end{abstract}

\date{\today}

\maketitle


\section{I. Introduction}

The physics of massless Dirac points, as exhibited in the band structure
of the honeycomb lattice of graphene, has driven intense
study\cite{castroneto09,geim09,choi10,novoselov12}.  The square lattice
with $\pi$-flux per plaquette is an alternate tight-binding Hamiltonian
which also contains Dirac points in its band structure.  Initial
investigations of the $\pi$-flux model focused on the non-interacting
limit\citep{harris89}, but, as with the honeycomb lattice, considerable
subsequent effort has gone into extending this understanding to
incorporate the effect of electron-electron interactions.  Numerical
simulations of the Hubbard Hamiltonian with an on-site repulsion $U$
between spin up and spin down fermions, including Exact Diagonalization
\citep{jia13} and Quantum Monte Carlo (QMC)\cite{otsuka02, otsuka14, li15,toldin15, otsuka16, lang18, guo18quantum,guo18uncon} revealed a quantum
phase transition at $U_{c} \sim 5.55\,t$ into a Mott antiferromagnetic
(AF) phase  in the chiral Heisenberg Gross-Neveu universality class.
For a spinless fermion system with near-neighbor interactions
a chiral Ising Gross-Neveu universality
class is suggested\cite{wang14}. These results have been contrasted with
those on a honeycomb lattice, which has a similar Dirac point structure,
though at a smaller critical interaction $U_{c} \sim 3.85\,t$
\cite{otsuka16}.

In the case of the repulsive Hubbard Hamiltonian, there were two
motivations for studying both the honeycomb and the $\pi$-flux
geometries.  The first was to verify that the quantum critical
transitions to AF order as the on-site repulsion $U$
increases share the same universality class, that of the Gross-Neveu
model.  The second was to confirm that an intermediate spin-liquid (SL)
phase between the semi-metal and AF phases\cite{meng10}, which had been
shown not to be present on a honeycomb lattice\cite{otsuka13}, was also
absent on the $\pi$-flux geometry.

Studies of the SU(2) $\pi$-flux Hubbard model have also been extended to
SU(4), using projector QMC\cite{zhou18}, and to staggered
flux where $\pm\pi$ hopping phases alternate
on the lattice\cite{chang12}.  In the former case, the semi-metal to AF
order transition was shown to be replaced by a semi-metal to valence
bond solid transition characterized by breaking of a ${\cal Z}_4$
symmetry.  In the latter work, an intermediate phase with power-law
decaying spin-spin correlations was suggested to exist between
the semi-metal and AF.

A largely open question is how this physics is affected in the presence of electron-phonon rather than electron-electron interactions.
A fundamental Hamiltonian, proposed by Holstein\citep{holstein59}, includes an 
on-site coupling of electron density to the linear displacement of the phonon field.  
In the low density limit, extensive numerical work has
quantified
polaron and bipolaron formation, in which electrons are ``dressed' by an accompanying lattice distortion
\cite{kornilovitch98,kornilovitch99,alexandrov00,hohenadler04,ku02,spencer05,macridin04,romero99}.
At sufficiently large coupling, electrons or pairs of electrons 
can become `self-trapped' (localized).
One of the most essential features of the Holstein model is that
the lattice distortion of one electron creates an energetically
favorable landscape for other electrons, so that there is an
effective attraction mediated by the phonons.
At higher densities, collective phenomena such as
Charge-Density Wave (CDW) phases,
and superconductivity (SC)
have been widely studied \citep{scalettar89, marsiglio90, vekic92, niyaz93, vekic93, Freericks93, zheng97, jeckelmann99, hohenadler04}.
CDW is especially favored on bipartite lattices and at fillings which
correspond to double occupation of one of the two sublattices.  SC
tends to occur when one dopes away from these commensurate fillings.

Recent work on the Holstein model
on the honeycomb lattice suggested a quantum phase transition from semi-metal to gapped CDW order \citep{zhang19,chen19} similar to the results for the Hubbard Hamiltonian.  However, a key difference between the Hubbard and Holstein models is the absence of the SU(2) symmetry of the order parameter in the latter case.  Thus, while long-range AF order arising from electron-electron interaction occurs only at zero temperature in 2D, the CDW phase transition induced by electron-phonon coupling can occur at finite temperature- the symmetry being broken is that associated with two {\it discrete} sub-lattices. For classical phonons ($\omega_0=0$), the electron-phonon coupling becomes an on-site energy in the mean-field approximation. In the anti-adiabatic limit where phonon frequencies are set to infinity, the Holstein model maps onto the attractive Hubbard model.

Here we extend the existing work on the effect of EPC on Dirac fermions
from the honeycomb geometry to the $\pi$-flux lattice.
The $\pi$-flux state is realized by threading half of a magnetic flux quantum through each plaquette of a square lattice\cite{affleck1988}.  
Recently it has been experimentally realized in optical lattices using Raman assisted
hopping\cite{aidelsburger2011}. There are also theoretical suggestions that the $\pi$-flux 
lattice might be engineered by the proximity of an Abrikosov lattice of vortices of a type-II superconductor, or via spontaneously generating a $\pi$-flux by coupling fermions to a ${\cal Z}_2$ gauge theory in (2+1) dimensions\cite{gazit2017}.
The $\pi$-flux hopping configuration
has an additional interesting feature motivating
our current work:  it is the unique magnetic field value which minimizes
the ground state energy for non-interacting fermions
at half-filled on a bipartite lattice.   Indeed, Lieb has shown that
this theorem is also true at finite temperature,
and furthermore holds in the presence of Hubbard
inteactions\cite{lieb94}.  Here we consider the thermodynamics of
the $\pi$-flux lattice with EPC.

This paper is organized as follows: in the next section, we describe
the Holstein model and the $\pi$-flux square lattice.
Section III presents, briefly, a mean-field theory (MFT) for the model.
Section IV reviews our primary method, DQMC. Section V contains results from the DQMC
simulations, detailing the nature of the CDW phase transition, both
the finite temperature transition at fixed EPC, and the QPT which occurs at $T=0$ with varying EPC.
Section VI contains our conclusions.


\section{II. Model}

The Holstein model \cite{holstein59} describes conduction electrons
locally coupled to phonon degrees of freedom,
\begin{align} \label{eq:Holst_hamil}
\mathcal{\hat H} = & - \sum_{\langle \mathbf{i}, \mathbf{j}
  \rangle, \sigma} \big(t_{\mathbf{i},\mathbf{j}}
\, \hat d^{\dagger}_{\mathbf{i} \sigma} \hat
d^{\phantom{\dagger}}_{\mathbf{j} \sigma} + {\rm h.c.} \big) - \mu
\sum_{\mathbf{i}, \sigma} \hat n_{\mathbf{i}, \sigma}
\nonumber \\
& + \frac{1}{2 M} \sum_{ \mathbf{i} } \hat{P}^{2}_{\mathbf{i}} +
\frac{\omega_{\, 0}^{2}}{2} \sum_{ \mathbf{i} }
\hat{X}^{2}_{\mathbf{i}} + \lambda \sum_{\mathbf{i}, \sigma} \hat
n_{\mathbf{i}, \sigma} \hat{X}_{\mathbf{i}} \,\,.
\end{align}
The sums on $\mathbf{i}$ and $\sigma$ run over all lattice sites and spins
$\sigma=\uparrow,\downarrow$.
$\langle \mathbf{i}, \mathbf{j} \rangle$ denotes
nearest neighbors. $\hat d^{\dagger}_{\mathbf{i} \sigma}$ and $\hat
d^{\phantom{\dagger}}_{\mathbf{i} \sigma}$ are creation and annihilation
operators of electrons with spin $\sigma$ on a given site $\mathbf{i}$;
$\hat n_{\mathbf{i}, \sigma}=\hat d^{\dagger}_{\mathbf{i} \sigma}\hat
d^{\phantom{\dagger}}_{\mathbf{i} \sigma}$ is the number operator. The
first term of Eq.~(1) corresponds to the hopping of electrons $\mathcal{K}_{\rm el}$, with
chemical potential $\mu$. The next line of the Hamiltonian describes optical phonons, local quantum harmonic oscillators of frequency
$\omega_{0}$ and phonon position and momentum operators,
$\hat{X}_{\mathbf{i}}$ and $\hat{P}_{\mathbf{i}}$ respectively.
The phonons are dispersionless since there are no terms connecting
$\hat X_{\bf i}$ on different sites of the lattice.
The phonon mass $M$ is set to unity.  The
electron-phonon coupling is included in the last term.
We set hopping  $|t_{\mathbf{i},\mathbf{j}}|=t=1$ as the energy
scale and
focus on half-filling,
($\langle \hat n \rangle=1$), which can be achieved
by setting $\mu=-\lambda^2/\omega_{0}^2$.  It is useful to present
results in terms of the
dimensionless coupling $\lambda_{D}=\lambda^2/(\omega_{0}^2 W)$
which represents the ratio of the effective electron-electron
interaction obtained after integrating out the phonon
degrees of freedom, and $W$ is the kinetic energy bandwidth.

\begin{figure}[t]
\includegraphics[height=2.40in,width=2.40in,angle=0]{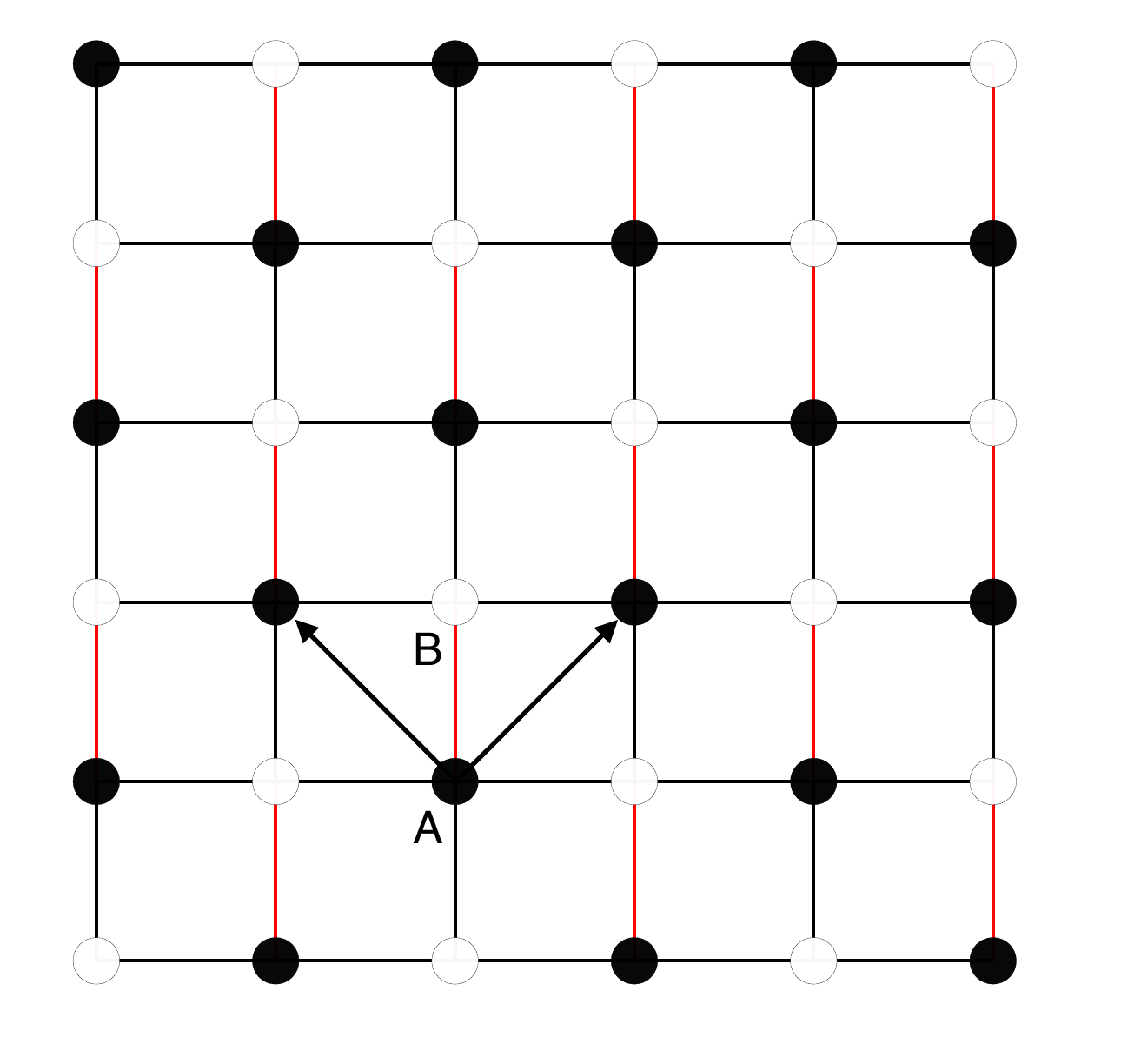} \caption{
$\pi$-flux phase on a $6 \times 6$ square lattice. Sublattices A and B
are shown by solid and open circles. Bonds in red correspond to hopping
$t'=-t$, as opposite to black lines with hopping $t$. Arrows represent the basis vectors.
}
\label{fig:pifluxgeom}
\end{figure}

\begin{figure}[t]
\includegraphics[height=2.40in,width=3.30in,angle=0]{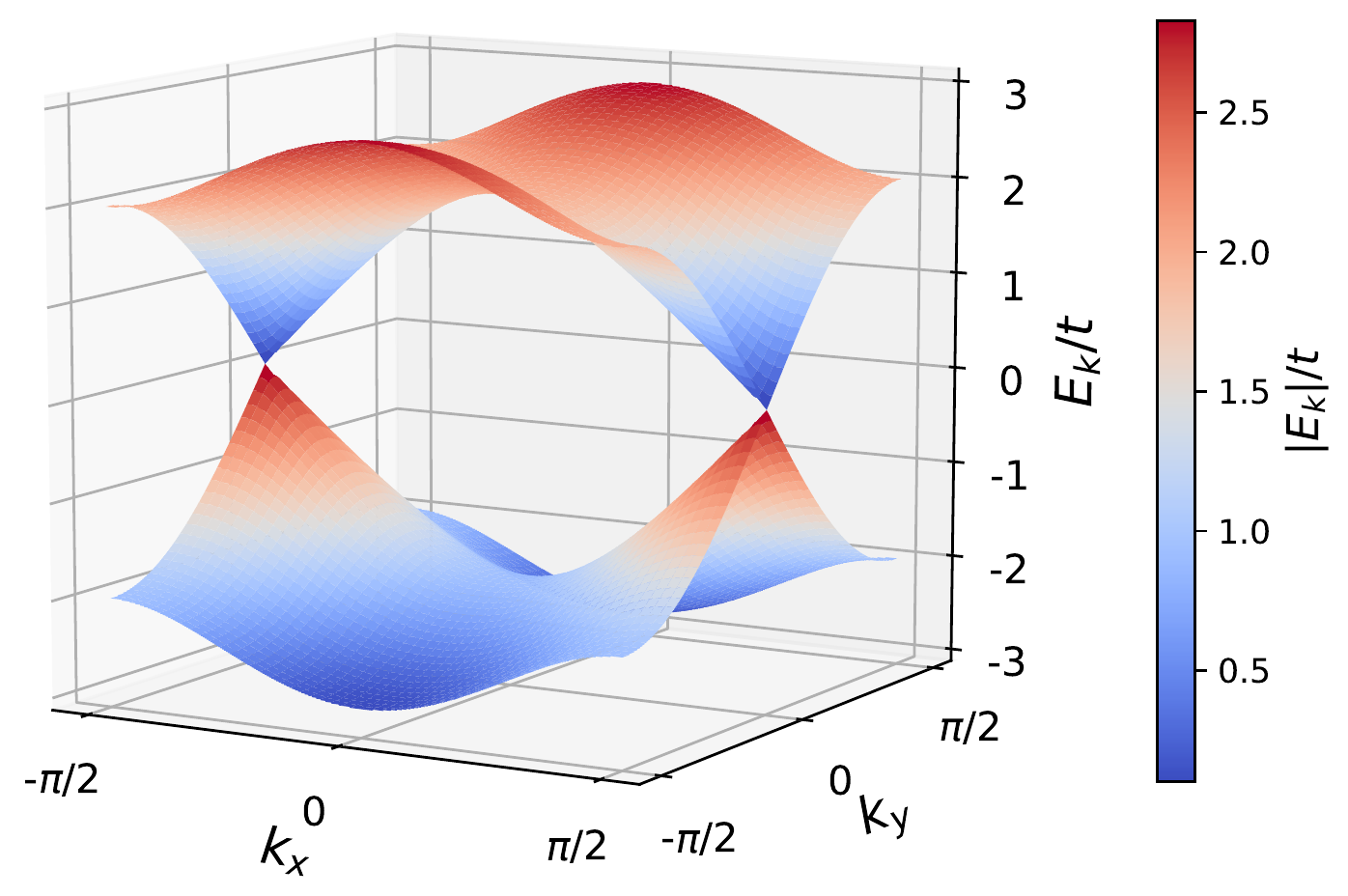}
\caption{The dispersion relation $E_{\bf k}$ for $\pi$-flux phase on a
square lattice.  There are two Dirac points at
$(k_x,k_y) = (\pm\pi/2,0)$. The bandwidth for the $\pi$-flux model is $W=4\sqrt{2}\,t$.
}
\label{fig:dispersion}
\end{figure}

\begin{figure}[t]
\includegraphics[scale=0.53]{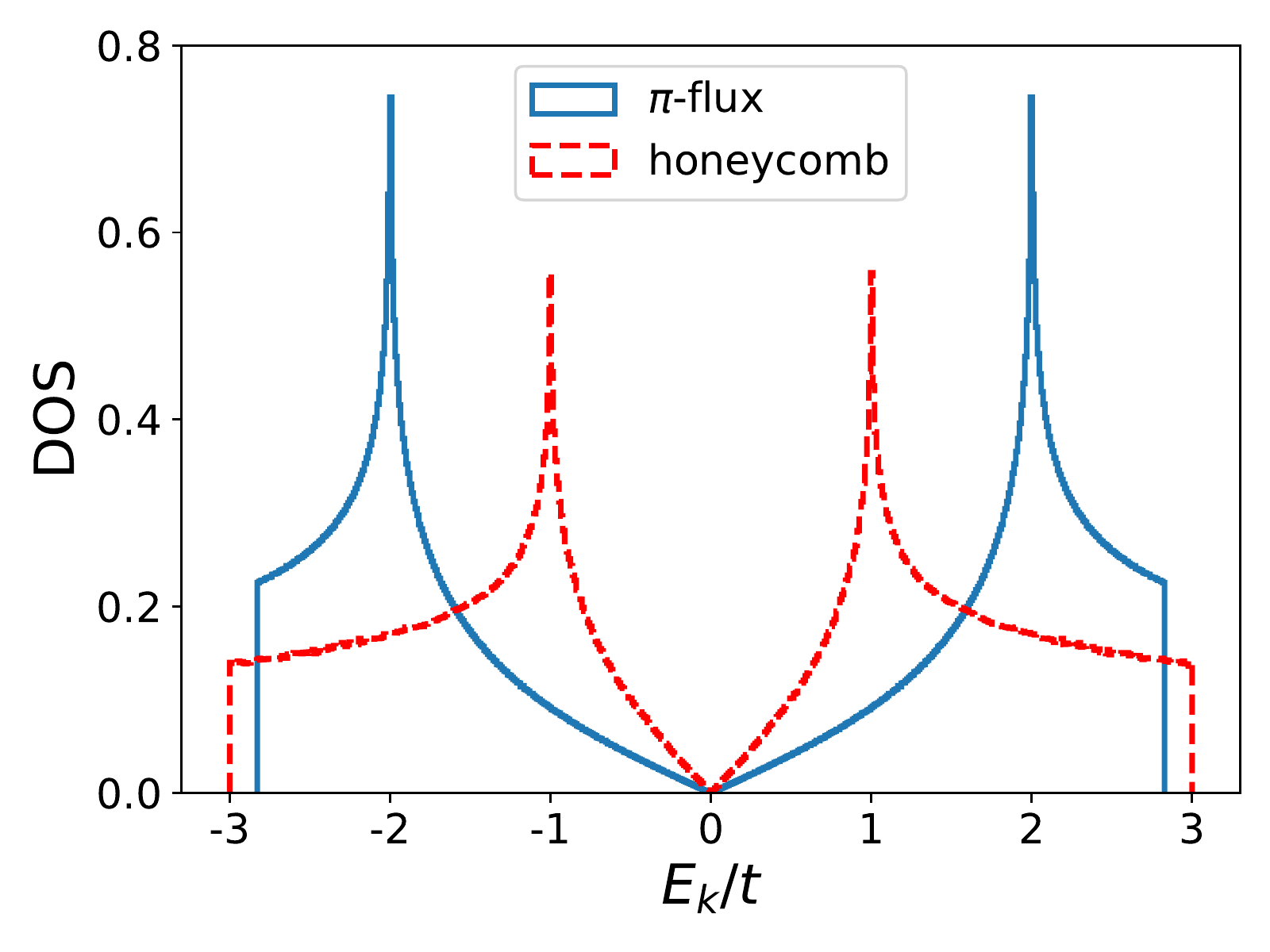}
\caption{ The density of states for the $\pi$-flux phase square
lattice and the honeycomb lattice.
The bandwidths are nearly identical, but the honeycomb lattice
has a substantially larger slope of the linear increase of the DOS.
}
\label{fig:dos}
\end{figure}

The two dimensional $\pi$-flux phase on a square lattice is
schematically shown in Fig.~\ref{fig:pifluxgeom}.  All hopping in the
$x$ direction are $t$, while half of the hoppings along the
$y$-direction are set to $t^\prime=t \, e^{i \pi}=-t$, where the phase $\pi$ in the hopping amplitude arises from the Peierls prescription for the vector potential of the magnetic field. As a consequence, an electron hopping on a contour around each plaquette picks up a total
phase $\pi$, corresponding to one half of a magnetic flux quantum $\Phi_{0}=hc/e$ per plaquette. The lattice is bipartite, with two
sublattices $A$ and $B$.  Each unit cell consists of two sites.  In
reciprocal space, with the reduced Brillouin zone $(|k_x| \leq \pi,
|k_y| \leq |k_x|)$, the non-interacting part of Hamiltonian Eq.(1) can
be written as,
\begin{align}
\mathcal{\hat H}_{0}  &=\sum_{\bf{k} \sigma}
 \hat \psi^{\dagger}_{\bf{k} \sigma}
{\bf H}_0({\bf k})
\hat \psi^{\phantom{\dagger}}_{\bf{k}\sigma} ,
\end{align}
where
\begin{align}
\hat \psi_{\bf{k}\sigma} = \left( \begin{array}{cc}
\hat d^{\phantom{\dagger}}_{A\sigma} &
\hat d^{\phantom{\dagger}}_{B\sigma} \\
\end{array} \right)^{T},
\end{align}
and the noninteracting Hamiltonian matrix
\begin{align}
 {\bf H}_0({\bf k}) =
\left( \begin{array}{cc}
 0 & 2\,t \, {\rm cos}k_x+2\, i \, t \, {\rm sin}k_y \\
 2\,t \, {\rm cos}k_x-2 \, i \, t \, {\rm sin}k_y & 0 \\
\end{array} \right).
\end{align}
The energy spectrum $E_{\bf k} = \pm 2\, t \sqrt{\cos^2 k_x+\sin^2
k_y}$ describes a semi-metal with two inequivalent Dirac points at
${\bf K}_{\pm}=(\pm\pi/2,0)$, shown in Fig.~\ref{fig:dispersion}.  In the
low-energy regime of the dispersion, the density of states (DOS)
vanishes linearly near the Dirac point where $E_{\bf k}=0$, as shown in Fig.~\ref{fig:dos}.
The bandwidth of the $\pi$-flux phase is $W=4\sqrt{2}\,t$.  In
Fig.~\ref{fig:dos} the DOS of the honeycomb lattice is shown for
comparison. The Dirac Fermi velocity is $v_{\rm F}=2t\ (1.5t)$ for the  $\pi$-flux (honeycomb) lattice. Near the Dirac point, the DOS
$\rho(\omega) \sim |\omega|/v_{\rm F}$, and the $\pi$-flux model has a smaller slope.

\section{III. Mean-Field Theory}

In this section, we present a
mean-field theory approach to solve the Holstein model.
Semi-metal to superfluid transitions have previously
been investigated with MFT in 2D and 3D
\cite{mazzucchi13,wu14}.
Here we focus on the
semi-metal to CDW transition.
In the mean-field approximation, the phonon displacement at site ${\bf
i}$ is replaced by its average value, modulated by a term
which has opposite sign on the two sublattices,
\begin{align}
\langle {X}_{\mathbf{i}}\rangle=X_{0} \pm X_{\rm mf} \, (-1)^{\bf i} \,\,.
\label{eq:MFansatz}
\end{align}
Here
$X_{0}=-\lambda/\omega_{0}^2$ is the ``equilibrium position" at
half-filling and $X_{\rm mf}$ is the mean-field order parameter. When $X_{\rm mf}$ = 0, phonons on all sites have the same average displacement, indicating the system remains in semi-metal phase, whereas when $X_{\rm mf} \neq 0$, the last term in the Hamiltonian Eq.~(1), i.e., $\lambda\sum_{{\bf i},\sigma}\hat{n}_{{\bf i},\sigma}\hat{X}_{\bf i}$, becomes an on-site staggered potential, which corresponds to the CDW phase.
The phonon kinetic energy term is zero as a result of the static field.
The resulting static mean-field Hamiltonian
is quadratic in the fermion operators.
Diagonalizing gives energy eigenvalues
$\epsilon_{n}(X_{\rm mf})$.
The free energy $F$ can then be directly obtained by,
\begin{align}
    F(\beta, X_{\rm mf})=- \frac{1}{\beta} \sum_{n} {\rm ln}(1+e^{-\beta
\epsilon_{n}})
\nonumber  \\
+ \frac{N \omega_{0}^2}{2} (X_{0}^2 + X_{\rm mf}^2),
\label{eq:MF}
\end{align}
Minimizing the free energy with respect to $X_{\rm mf}$
(or equivalently, a self-consistent calculation)
will determine the order parameter.
$X_{\rm mf}$ is found to be zero at high temperatures:
the energy cost of the second term in Eq.~\ref{eq:MF} exceeds
the energy decrease in the first term associated with opening of a gap
in the spectrum $\epsilon_n$.
$X_{\rm mf}$ becomes nonzero below a critical temperature $T_{c}$.

\begin{figure}[t]
\includegraphics[scale=0.53]{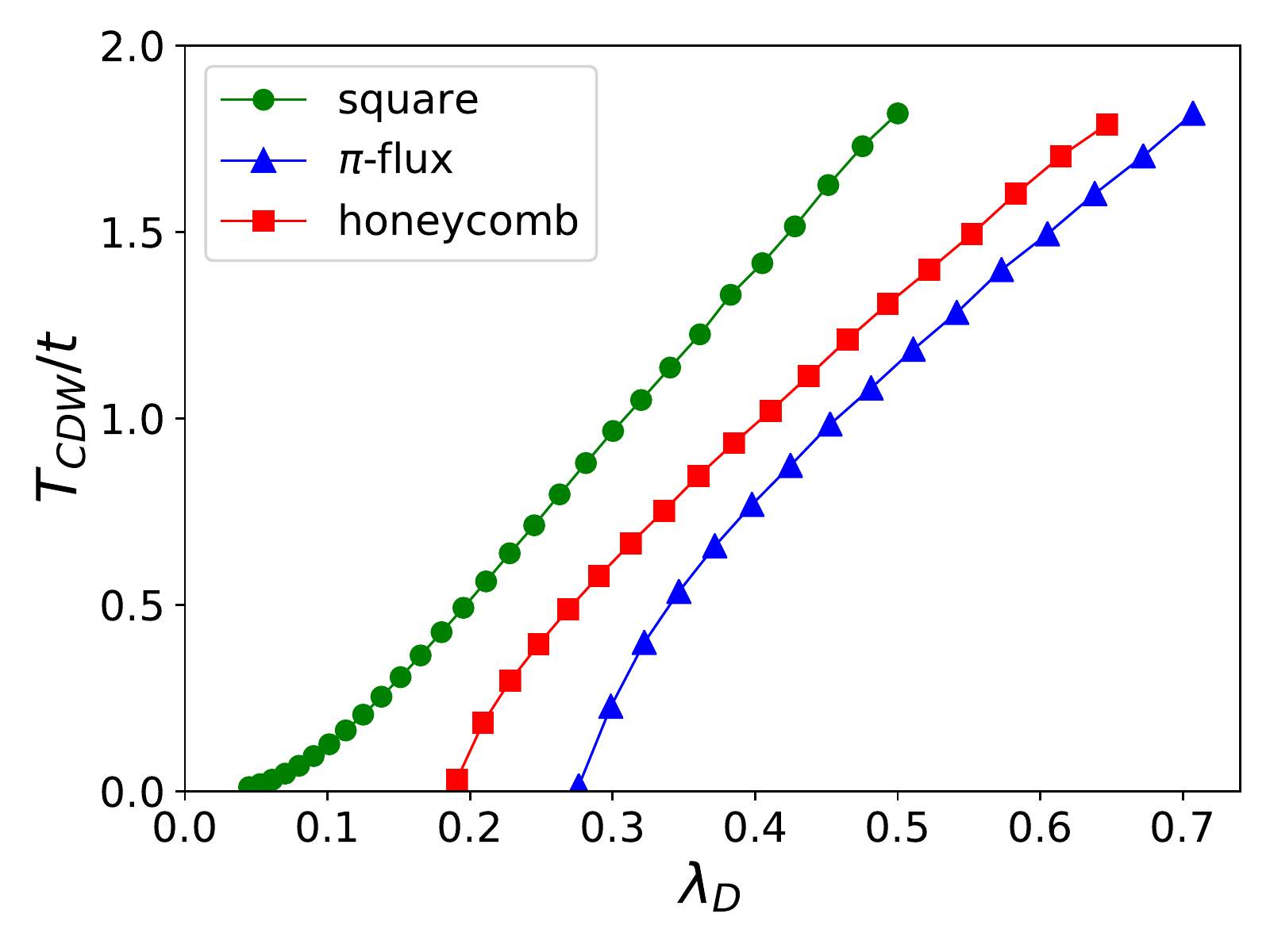}
\caption{MFT $T_{c}$ for CDW phase transition as a function of dimensionless coupling $\lambda_{D}$ for the square lattice
with no magnetic flux, the $\pi$-flux phase square lattice, and the
honeycomb lattice.  For the geometries with a Dirac spectrum MFT
captures the existence for a QCP, a critical value of $\lambda_D$ below
which there is no CDW order even at $T=0$, and the absence of a QCP for
the conventional square lattice.
}
\label{fig:MFT}
\end{figure}

$T_{c}$ for the $\pi$-flux lattice is shown in Fig.~\ref{fig:MFT},
along with the result of analogous MFT calculations
for the honeycomb and (zero flux)
square geometries.
The lattice size $\rm L = 180$ is chosen for all three models. This is sufficiently large so that finite size effects are smaller than the statistical sampling error bars.
 At zero temperature, the CDW order
exhibits a critical EPC for the $\pi$-flux
and the honeycomb lattices.  This QCP
arises from the Dirac fermion dispersion, which has a vanishing DOS
at the Fermi energy.
The honeycomb lattice QCP has a smaller critical value.
However, when measured in units of the Fermi velocity,
the ratios
$\lambda_{D,{\rm crit}}/v_{\rm F}=0.13$ and $0.14$
are quite close
for the honeycomb and $\pi$-flux geometries respectively.
We will see this is also the case for the exact DQMC calculations.
For the square lattice, on the other hand, the DOS has a Van-Hove
singularity at the Fermi energy, and the CDW develops at arbitrarily small
coupling strength.

Another feature of the MFT phase diagram is that,
as the coupling increases,
$T_{c}$ increases monotonically.  This is in contrast to the
exact DQMC results, where $T_c$ decreases at large coupling strengths
(Fig.~\ref{fig:PD}).
A similar failure of MFT is well known for the Hubbard Hamiltonian
where the formation of AF ordering is related to two factors: the local moment $m^2_z=(n_\uparrow-n_\downarrow)^2=1-2\langle n_\uparrow n_\downarrow\rangle$ and the exchange coupling $J \sim t^2/U$. The double occupancy $\langle n_\uparrow n_\downarrow\rangle$ is suppressed by the interaction, resulting in the growth of the local moment. Thus upon cooling, the Hubbard model has two characteristic temperatures: the temperature of local moment formation, which increases monotonically with $U$, and further the AF ordering scale, which falls as $J$. Since the interaction is simply decoupled locally and the exchange coupling is not addressed, within MFT the formation of the local 
moments, and their ordering, occur simultaneously.
MFT thus predicts a monotonically increasing $T_c$ with $U$.


\section{IV.  DQMC Methodology}

We next describe the DQMC
method\cite{blankenbecler81,white89}.
In evaluating the partition function $\mathcal{Z}$,
the inverse temperature $\beta$ is discretized as $\beta=L_{\tau}
\Delta \tau$, and complete sets of phonon
position eigenstates are introduced between each
$e^{-\Delta \tau {\cal \hat H}}$.
The phonon coordinates acquire an ``imaginary time" index,
converting the 2-dimensional quantum system to a (2+1) dimensional
classical problem.
After tracing out the fermion degrees of freedom, which appear only
quadratically in the Holstein Hamiltonian, the partition function becomes
\begin{align}
\mathcal{Z} =  \int \mathcal{D}x_{\mathbf{i},l} \, e^{-\mathcal{S}_{ph}} \left[ \det{\mathbf{M} (x_{\mathbf{i},l})} \right]^2 ,
\label{eq:Z}
\end{align}
where the ``phonon action" is
\begin{align}
\mathcal{S}_{ph}=\Delta \tau \left[\frac{1}{2}\omega_{0}^2
\sum_{\mathbf{i}}x_{\mathbf{i},l}^2 + \frac{1}{2 M}\sum_{\mathbf{i}}
\left(\frac{x_{\mathbf{i},l+1}-x_{\mathbf{i},l}}{\Delta \tau} \right)^2
\right] \,.
\label{eq:Sbose}
\end{align}

Because the spin up and spin down fermions have an identical
coupling to the phonon field, the fermion determinants which result
from the trace are the same, and the determinant
is squared in Eq.~\ref{eq:Z}.
Thus there is no fermion sign problem\cite{loh90}.
We use $\Delta \tau=0.1/t$, small enough so that Trotter
errors associated with the discretization of $\beta$ are of the same
order of magnitude as the statistical uncertainty from the Monte Carlo
sampling.

\begin{figure}[t]
\includegraphics[scale=0.44]{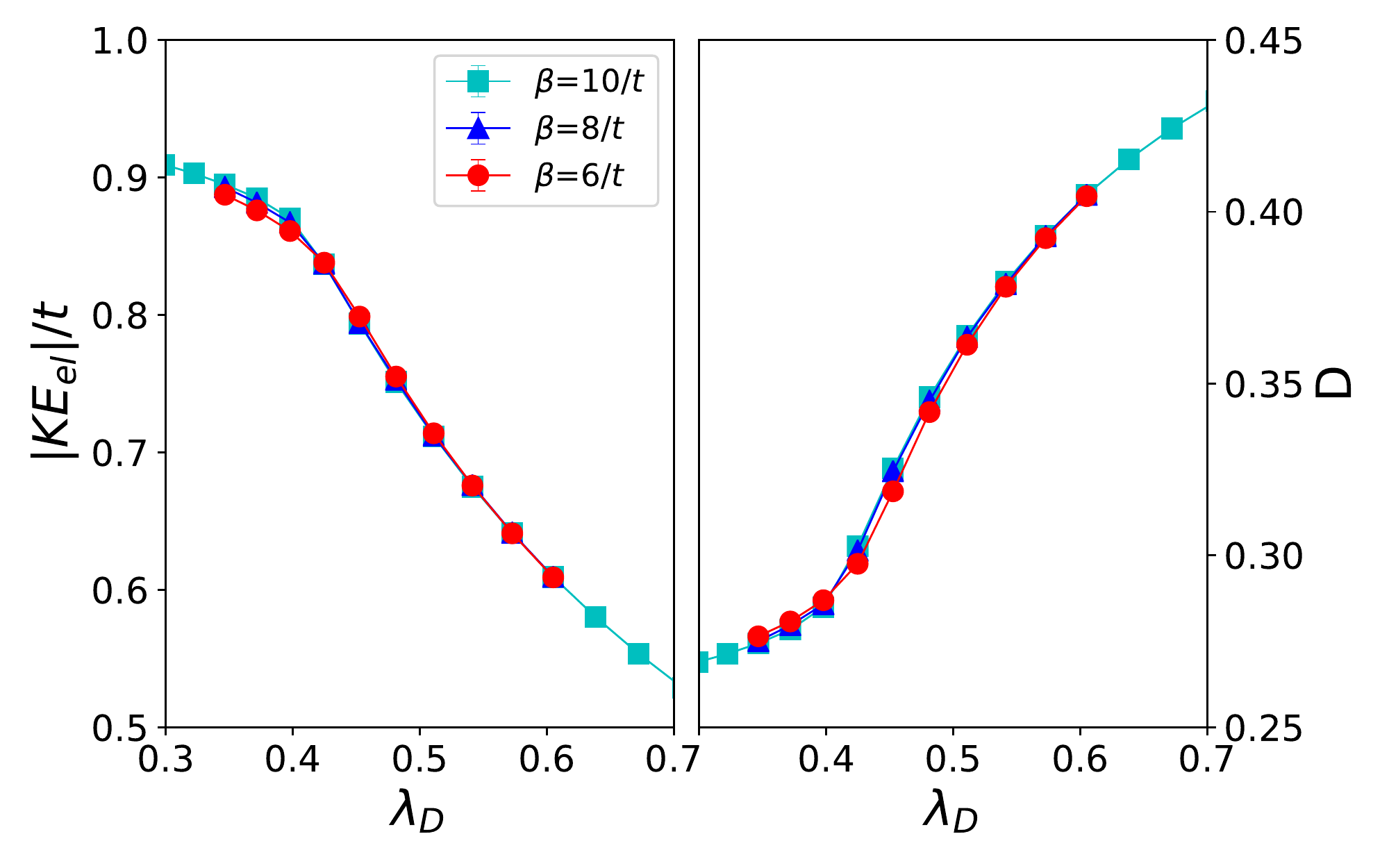}
\caption{
\underbar{Left:}
The magnitude of electron kinetic energy $|\mathcal{K}_{\rm el}|$
as a function of EPC strength $\lambda_{D}$. Simulations are performed
on a $\rm L=10$ lattice at inverse temperatures $\beta=6/t, 8/t, 10/t$ and fixed $\omega_0=1.0 \,t$.
\underbar{Right:}
Double occupancy $D$ as a function of EPC strength
$\lambda_{D}$.
}
\label{fig:KEandD}
\end{figure}

\section{V.  DQMC Results}

\subsection{Double occupancy and Kinetic Energy}

We first show data for several local observables,
the electron kinetic energy
$|\mathcal{K}_{\rm el}|=|\sum_{\langle \mathbf{i}, \mathbf{j}
  \rangle, \sigma} \big(t^{\phantom{\dagger}}_{\mathbf{i},\mathbf{j}}
\, \hat d^{\dagger}_{\mathbf{i} \sigma} \hat
d^{\phantom{\dagger}}_{\mathbf{j} \sigma}+ {\rm h.c.} \big)|$ and double occupancy
$\mathcal{D} =
\langle \, n_{{\bf i}\uparrow} n_{{\bf i}\downarrow} \, \rangle$.
For a tight-binding model on a bipartite
lattice at half-filling, Lieb has shown that
the energy-minimizing magnetic flux is $\pi$ per plaquette,
both in for noninteracting fermions and in the presence of
a Hubbard $U$ \cite{lieb94}.
Here we show $|\mathcal{K}_{\rm el}|$ for the Holstein model,
a case not hitherto considered.

Figure~\ref{fig:KEandD} shows $|\mathcal{K}_{\rm el}|$ (left panel) and
$\mathcal{D}$ (right panel) as functions of the dimensionless EPC
$\lambda_D$ for $\beta=6/t,8/t,10/t$.  There is little temperature dependence for these local quantities.  The magnitude of the kinetic energy
$|\mathcal{K}_{\rm el}|$ decreases as $\lambda_D$ grows, reflecting the
gradual localization of the dressed electrons (``polarons").

At the same time, the double occupancy $\mathcal{D}$ evolves from its
noninteracting value $\mathcal{D} = \langle \, n_{{\bf i}\uparrow}
n_{{\bf i}\downarrow} \, \rangle = \langle \, n_{{\bf i}\uparrow} \,
\rangle \, \langle \, n_{{\bf i}\downarrow} \, \rangle = 1/4$ at
half-filling, to $\mathcal{D} = 1/2$ at large $\lambda_D$.  In the strong
coupling regime, we expect robust pair formation, so that half of the
lattice sites will be empty and half will be doubly occupied.

The evolution of $\mathcal{D}$  and $|\mathcal{K}_{\rm
el}|$ have largest slope at $\lambda_{D} \sim 0.42$ which,
as will be seen, coincides with the
location of the QCP between the semi-metal and
CDW phases.

\begin{figure}[t]
\includegraphics[height=3.00in,width=3.50in,angle=0]{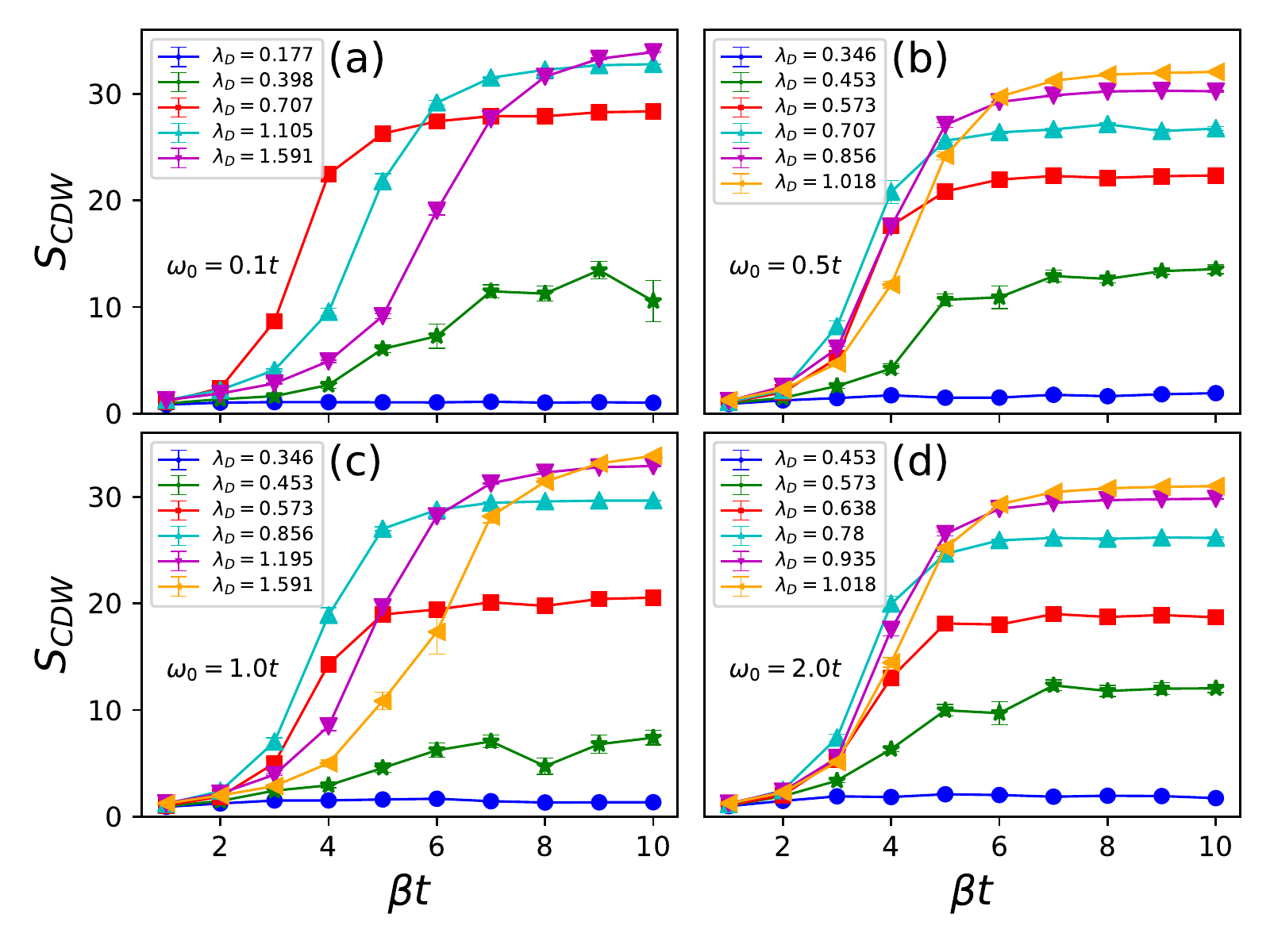}
\caption{ The CDW structure factor
of the $\pi$-flux phase Holstein model
as a function of inverse temperature
$\beta$. The phonon frequencies $\omega_{0}$ are (a), $0.1 \,t$; (b),
$0.5 \,t$; (c), $1.0 \,t$; (d), $2.0 \,t$ in the four panels.
The lattice size $L=6$.
}
\label{fig:Scdwvsbeta}
\end{figure}


\subsection{Existence of Long-Range CDW Order}

The structure factor $S({\bf Q})$
is the Fourier transform of the real-space spin-spin correlation
function $c({\bf r})$,
\begin{align}
S(\bf Q) &=  \sum_{\bf r} e^{i \bf Q \cdot \bf r} c({\bf r}),
\nonumber \\
c({\bf r}) &= \big\langle \,
\big( \, n_{{\bf i}\uparrow} + n_{{\bf i}\downarrow} \, \big)
\big( \, n_{{\bf i+r}\uparrow} + n_{{\bf i+r}\downarrow} \, \big)
\, \big\rangle,
\end{align}
and characterizes the charge ordering.  In a disordered phase $c({\bf
r})$ is short-ranged and $S({\bf Q})$ is independent of lattice size.
In an ordered phase, $c({\bf r})$ remains large out to long distances,
and the structure factor will be proportional to the number of sites, at the
appropriate ordering wave vector ${\bf Q}$.  At half-filling $S({\bf
Q})$ is largest at ${\bf Q}=(\pi,\pi)$.  We define $S_{\rm cdw} \equiv
S(\pi,\pi)$. Figure \ref{fig:Scdwvsbeta} displays $S_{\rm cdw}$ as a
function of inverse temperature $\beta$ at different phonon frequencies
$\omega_{0}$ and coupling strengths $\lambda_{D}$.  The linear lattice
size $L=6$. At fixed $\omega_{0}$ and strong coupling, $S_{\rm cdw}$
grows as temperature is lowered, and saturates to $S_{\rm cdw} \sim \rm
N$, indicating the development of long-range order (LRO), i.e.~the phase
transition into CDW phase. Note that $\beta=10/t$ is always in the plateau region, suggesting the correlation length has become larger than the lattice size, and the ground state has been reached.  In the following, we use $\beta=10/t$ to represent the properties at $T \rightarrow 0$.

However, as $\lambda_{D}$ is decreased sufficiently, $S_{\rm cdw}$
eventually shows no signal of LRO even at large $\beta$, providing an
indication that there is a QCP, with CDW order only occurring above a
finite $\lambda_{D}$ value.  Figure \ref{fig:Scdwvsbeta} also suggests
that the critical temperature $T_{c}$ is non-monotonic with increasing
$\lambda_{D}$.  The values of $\beta$ at which $S_{\rm cdw}$ grows first
shift downward, but then become larger again.  This non-monotonicity
agrees with previous studies of Dirac fermions on the honeycomb lattice
\cite{zhang19,chen19}.  We can estimate the maximum $T_{c}$ to occur at
$\lambda_{D} \approx 0.71, 0.71, 0.86$ and $0.78$ for $\omega_0 = 0.1 \,t, 0.5 \,t, 1.0 \,t, 2.0 \,t$ respectively.  In the anti-adiabatic limit $\omega_0
\rightarrow \infty$, the Holstein model maps onto the attractive Hubbard
model, and $T_c=0$ owing to the degeneracy of CDW and superconducting
correlations\cite{scalettar89}.  (The order parameter has a continuous
symmetry.) A recent study\cite{feng20} has shown that
$\omega_0 \gtrsim 10^2 \, t$ is required to achieve the $-U$ Hubbard model
limit, a surprisingly large value.

\begin{figure}[t]
\includegraphics[height=2.2in,width=3.4in,angle=0]{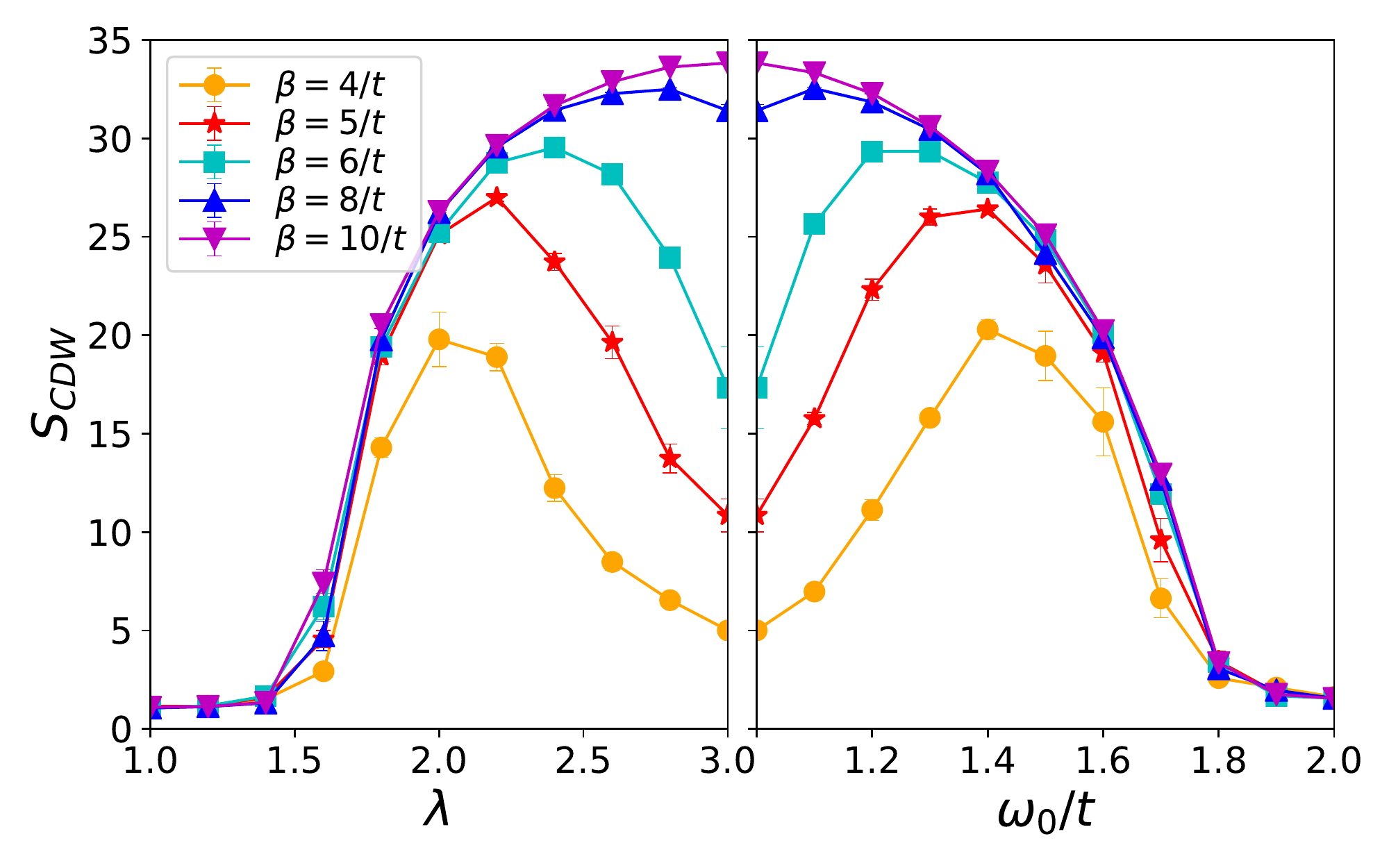}
\caption{$S_{\rm cdw}$ (a) as a function of $\lambda$ at fixed
$\omega_{0}=1.0 \,t$; and (b) as a function of $\omega_{0}$ at fixed
$\lambda$=3.0, at different inverse temperatures $\beta$. Lattice size $L=6$ is used in this figure.
}
\label{fig:Scdwvslamw}
\end{figure}

Figure \ref{fig:Scdwvslamw}(a) shows
$S_{\rm cdw}$ as a function of $\lambda$ at fixed $\omega_0=1.0 \,t$.
At the highest temperature shown, $\beta=4/t$, $S_{\rm cdw}$ reaches
maximum at intermediate coupling $\lambda \sim 2.0$, then decreases as
$\lambda$ gets larger.  The region for which $S_{\rm cdw}$ is large is a
measure of the range of $\lambda$ for which the CDW ordering temperature
$T_c$ exceeds $\beta^{-1}$.  As $\beta$ increases, this range is
enlarged.
Figure \ref{fig:Scdwvslamw}(b) is an analogous plot of $S_{\rm cdw}$ as
a function of $\omega_{0}$ at fixed $\lambda=3.0$.  The two plots appear
as mirror images of each other since the dimensionless EPC
$\lambda_D=\lambda^2/(\omega_0^2 W)$ increases with $\lambda$, but
decreases with $\omega_0$.

\begin{figure}[t]
\includegraphics[height=2.2in,width=3.45in,angle=0]{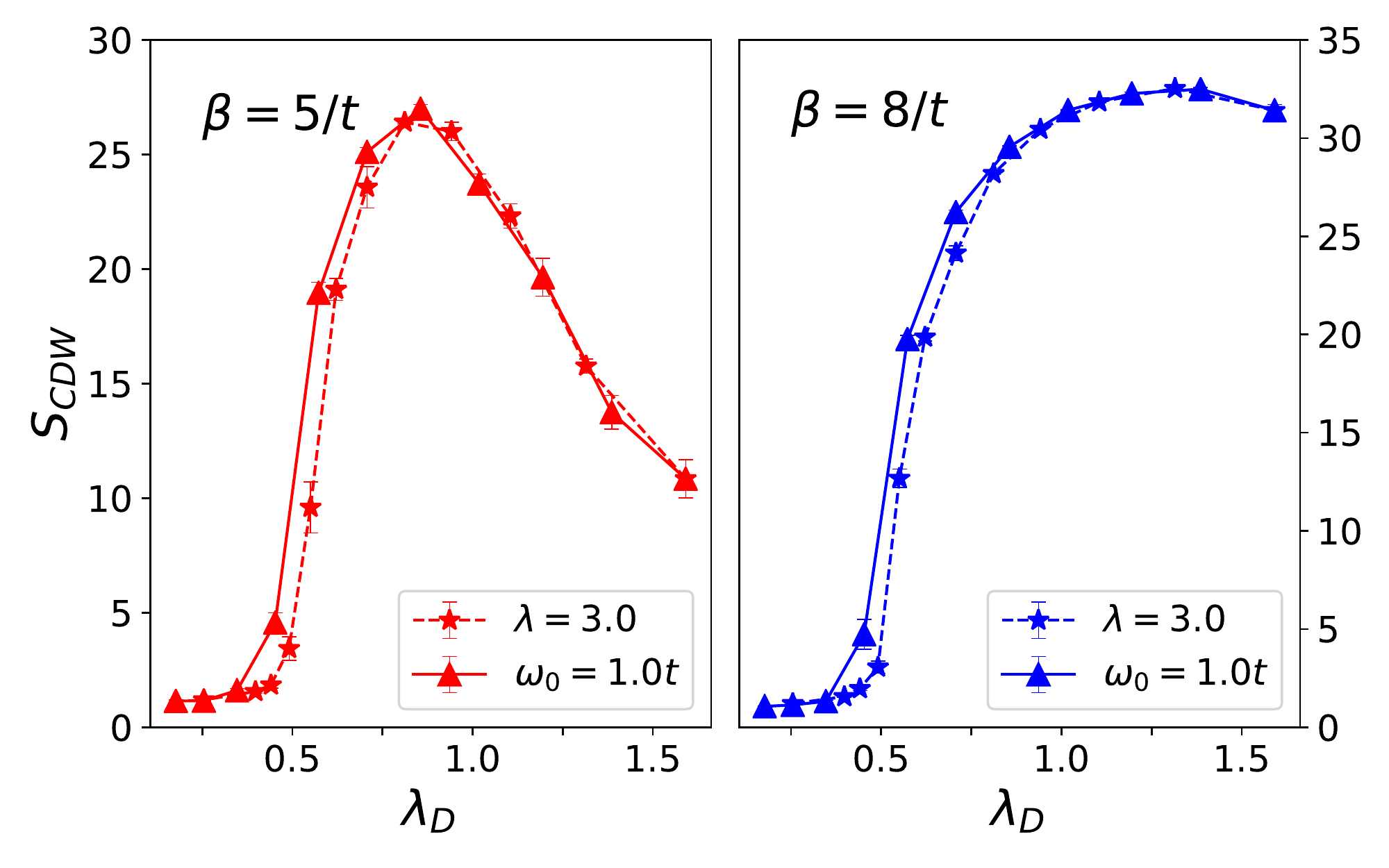}
\caption{Comparison of the evolution of
$S_{\rm cdw}$ with coupling strength by changing $\lambda$
or changing $\omega_{0}$. Data are taken from
Fig.~\ref{fig:Scdwvslamw}(a,b),
for $\beta=5/t$ (left) and $\beta=8/t$ (right).  The difference is
negligible at $\lambda_{D}>0.8$ but not in the coupling
regime $0.4<\lambda_{D}<0.8$ near the QCP. 
}
\label{fig:ScdwvslamD}
\end{figure}

It is interesting to ascertain the extent to which the physics of the
Holstein Hamiltonian is determined by $\lambda$ and $\omega_0$
separately, versus only the combination $\lambda_D$.  Figure
\ref{fig:ScdwvslamD} addresses this issue by replotting the data of
Figs.~\ref{fig:Scdwvslamw}(a,b) as a function of $\lambda_D$ for two
values of the inverse temperature.  For $\lambda_{D} \gtrsim 0.8$, the
data collapse well, whereas at small $\lambda_{D}$ $S_{\rm cdw}$ can
vary by as much as a factor of two even though $\lambda_D$ is identical.
It is likely that this sensitivity to the individual values of $\lambda$ and $\omega_0$ is associated with proximity to the QCP.

\begin{figure}[t]
\includegraphics[scale=0.55]{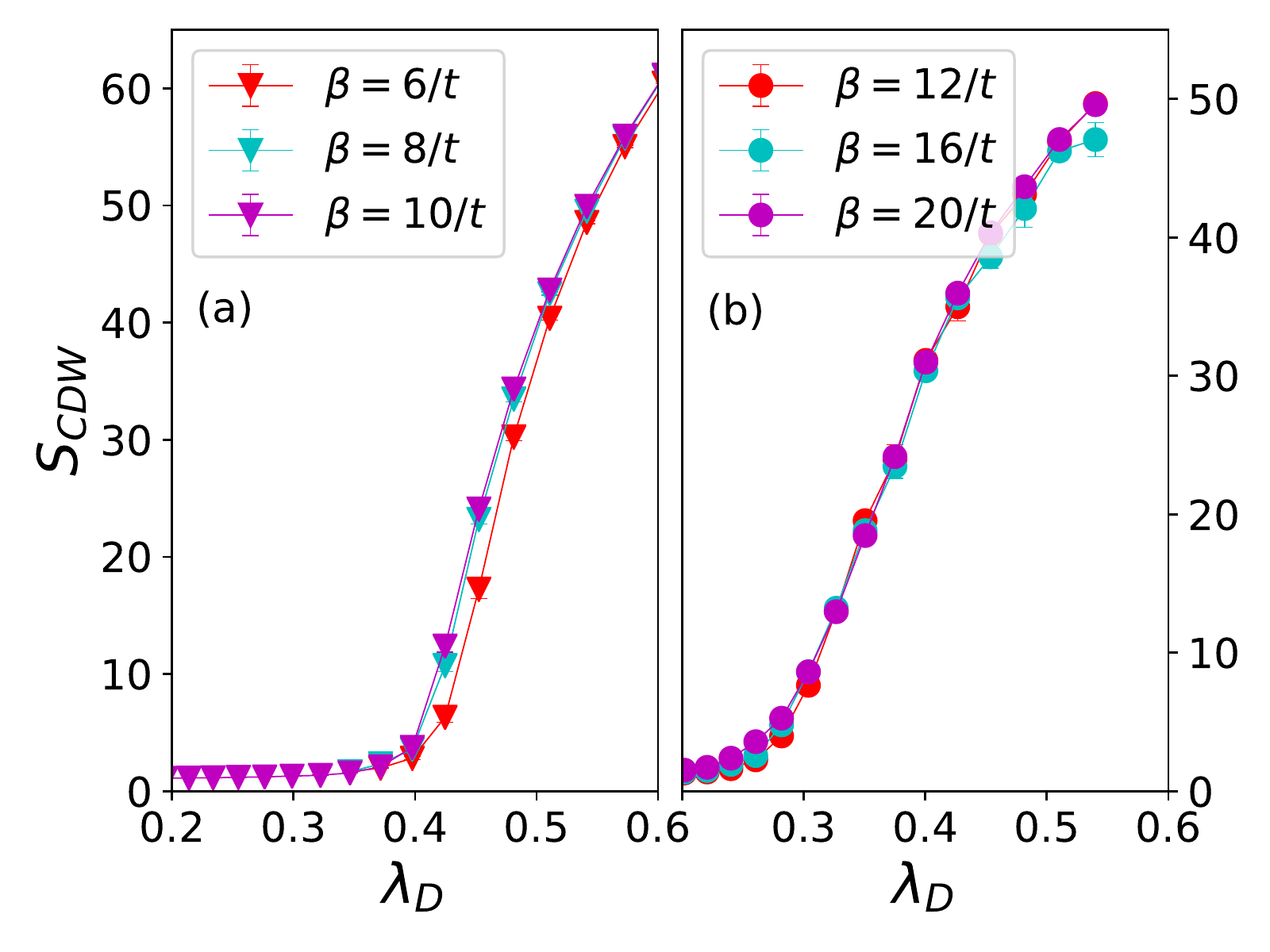}
\caption{$S_{\rm cdw}$ as a function of $\lambda_{D}$ for $\pi$-flux
phase square lattice (left) and honeycomb model (right).
The lattice size $L=6$ is used for both geometries.
$\lambda_D$ is varied by changing $\lambda$ at fixed $\omega_0=1.0 \,t$.
$S_{\rm cdw}$ does not change
for the lowest temperatures, indicating that the
ground state has been reached for this finite lattice size.
}
\label{fig:PIvsHC}
\end{figure}

We compare the semi-metal to CDW transition with increasing
$\lambda_D$ for the $\pi$-flux phase
and honeycomb lattices in Fig.~\ref{fig:PIvsHC}.
These data are at lower temperatures than those of
Fig.~\ref{fig:ScdwvslamD}, so that the ground state
values of $S_{\rm cdw}$ have been reached for the system sizes shown.



\subsection{Ground State in the ($\lambda,  \omega_{0}$) Plane}
\begin{figure}[t]
\includegraphics[scale=0.65]{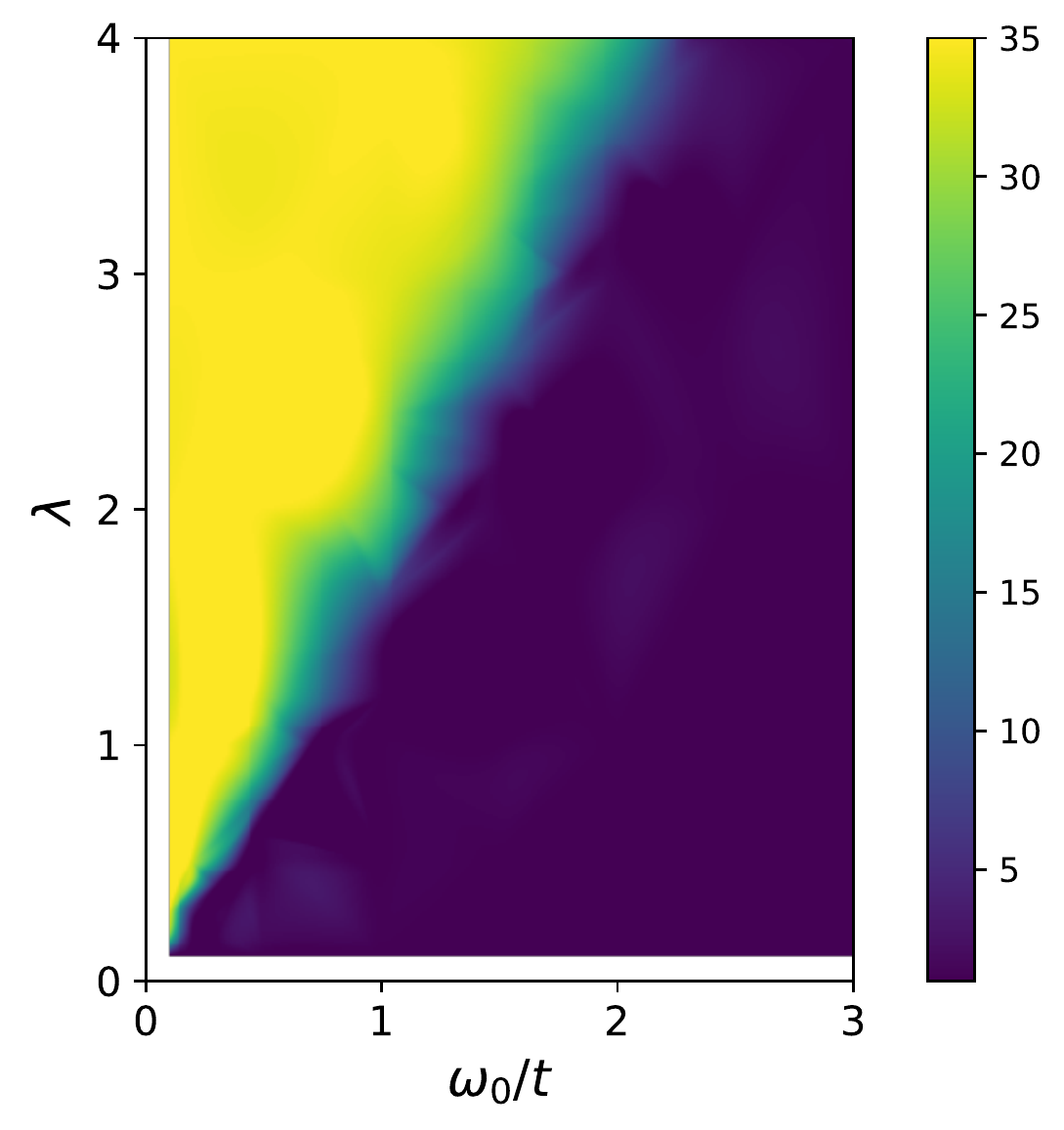}
\caption{Heat map of the ground state values of
$S_{\rm cdw}$ in the ($\lambda, \omega_{0}$) plane.
}
\label{fig:PDTeq0}
\end{figure}

Figure \ref{fig:PDTeq0} provides another perspective on the
dependence of the CDW order on $\lambda$ and $\omega_0$ individually,
by giving a heat map of $S_{\rm cdw}$ in the
($\lambda, \omega_{0}$) plane
at low temperature.
The bright yellow in upper-left indicates a strong CDW
phase, whereas the dark purple region in lower-right indicates the
Dirac semi-metal phase. The phase boundary
is roughly linear, as would be
expected if only the combination $\lambda_D = \lambda^2/(\omega_0^2 W)$
is relevant.
We note, however, that this statement is only qualitatively true.
The more precise line graphs of
Fig.~\ref{fig:ScdwvslamD} indicate that along the line
$\lambda=\sqrt{\lambda_{D,{\rm crit}}W} \, \omega_0 \sim 1.5 \,
\omega_0$, the separate values of $\lambda$ and $\omega_0$ are relevant.

\subsection{Finite Size Scaling: Finite $T$ Transition}

\begin{figure}[t]
\includegraphics[scale=0.59]{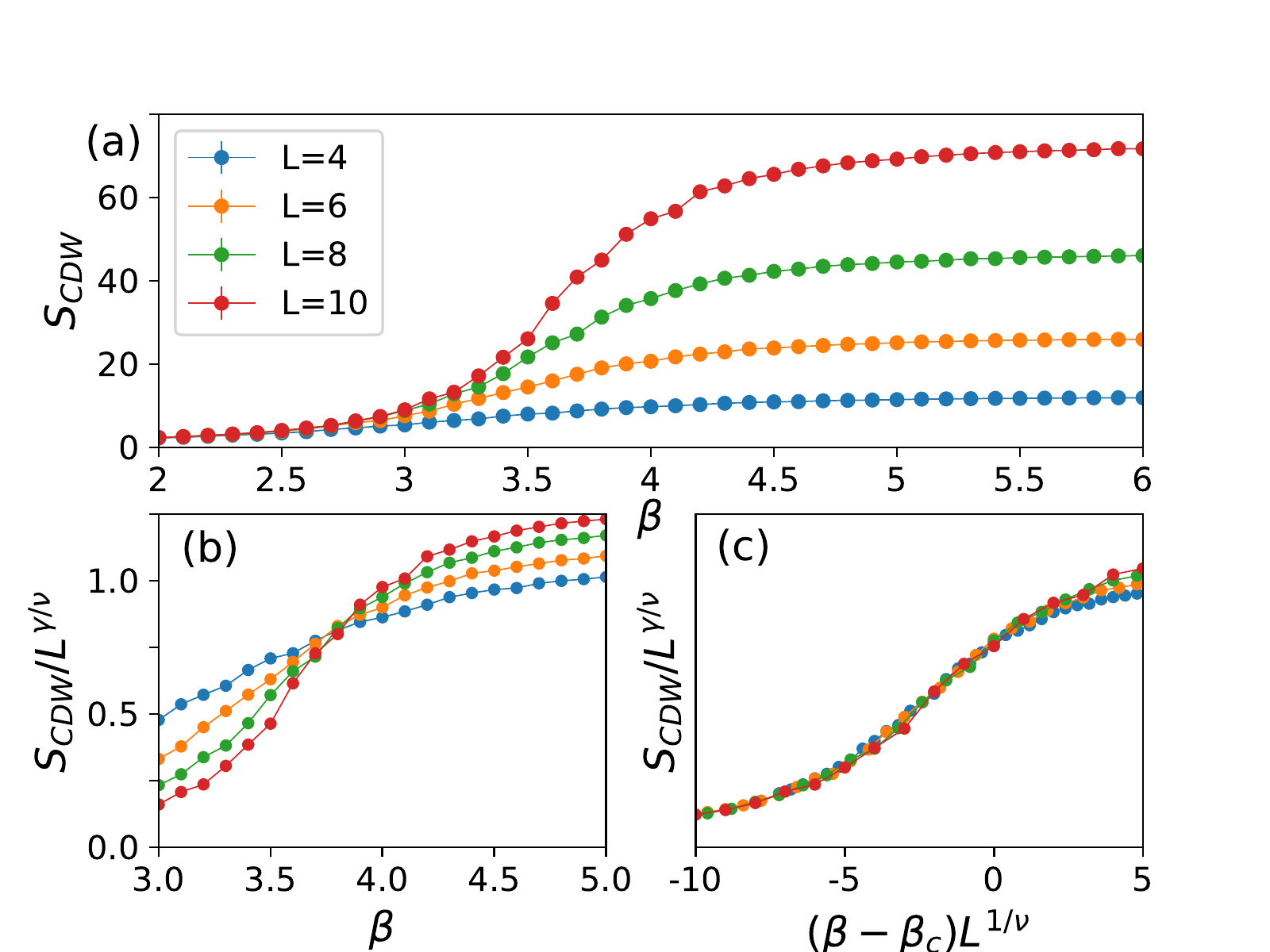}
\caption{(a) The CDW structure factor $S_{\rm cdw}$ as a function of
$\beta$ for several lattice sizes. (b) The scaled CDW structure factor
$S_{\rm cdw}/\mathrm{L}^{\gamma/\nu}$ as a function of $\beta$ using
Ising critical exponents $\gamma=7/4$ and $\nu=1$, showing a crossing of
different L at $\beta_{c}=3.80/t$. (c) $S_{\rm
CDW}/\mathrm{L}^{\gamma/\nu}$ versus $(\beta-\beta_{c}) \rm L$, giving a
best data collapse at $\beta_{c}=3.80/t$. Here the parameters are
$\lambda=2.0$ and $\omega_{0} =1.0\,t$.
}
\label{fig:FSS} %
\end{figure}

A quantitative determination of the finite temperature and quantum
critical points can be done with finite size scaling (FSS).
Figure \ref{fig:FSS} gives both raw and scaled data for $S_{\rm cdw}$
for different lattice sizes $L=4, 6, 8, 10$ at $\lambda=2.0$,
$\omega_{0}=1.0 \,t$ as a function of $\beta$.
Unscaled data are in panel (a):
$S_{\rm cdw}$ is small and $L$-independent at small $\beta$ (high $T$)
where $c({\bf r})$ is short ranged.
On the other hand,
$S_{\rm cdw}$ is proportional to $N=L^2$ at large $\beta$ (low $T$),
reflecting the long-range CDW order in $c({\bf r})$.
Panel (b) shows a data crossing for different $L$ occurs when
$S_{\rm cdw}/L^{\gamma/\nu}$ is plotted  versus $\beta$.
A universal crossing is seen at $\beta \,t \sim 3.80 \pm 0.02$, giving a precise determination of
critical temperature $T_{c}$.  The 2D Ising critical exponents $\gamma=7/4$ and
$\nu=1$ were used in this analysis, since the CDW phase transition breaks
a similar discrete symmetry.  Panel (c) shows a full data collapse when
the $\beta$ axis is also appropriately scaled by $L^{1/\nu}$.
The best collapse occurs at
$\beta_{c}=3.80/t$, consistent with the result from the data crossing.

In the region immeditely above the QCP,
the DQMC values for $T_c$ are roughly five times lower than
those obtained in MFT, and, indeed, the MFT over-estimation
of $T_c$ can be made arbitrarily large at strong
coupling.  This reflects both the relatively
low dimensionality ($d=2$) and the fact that
MFT fails to distinguish moment-forming and moment-ordering
temperature scales.

\subsection{Quantum Phase Transition}

\begin{figure}[t]
\includegraphics[scale=0.55]
{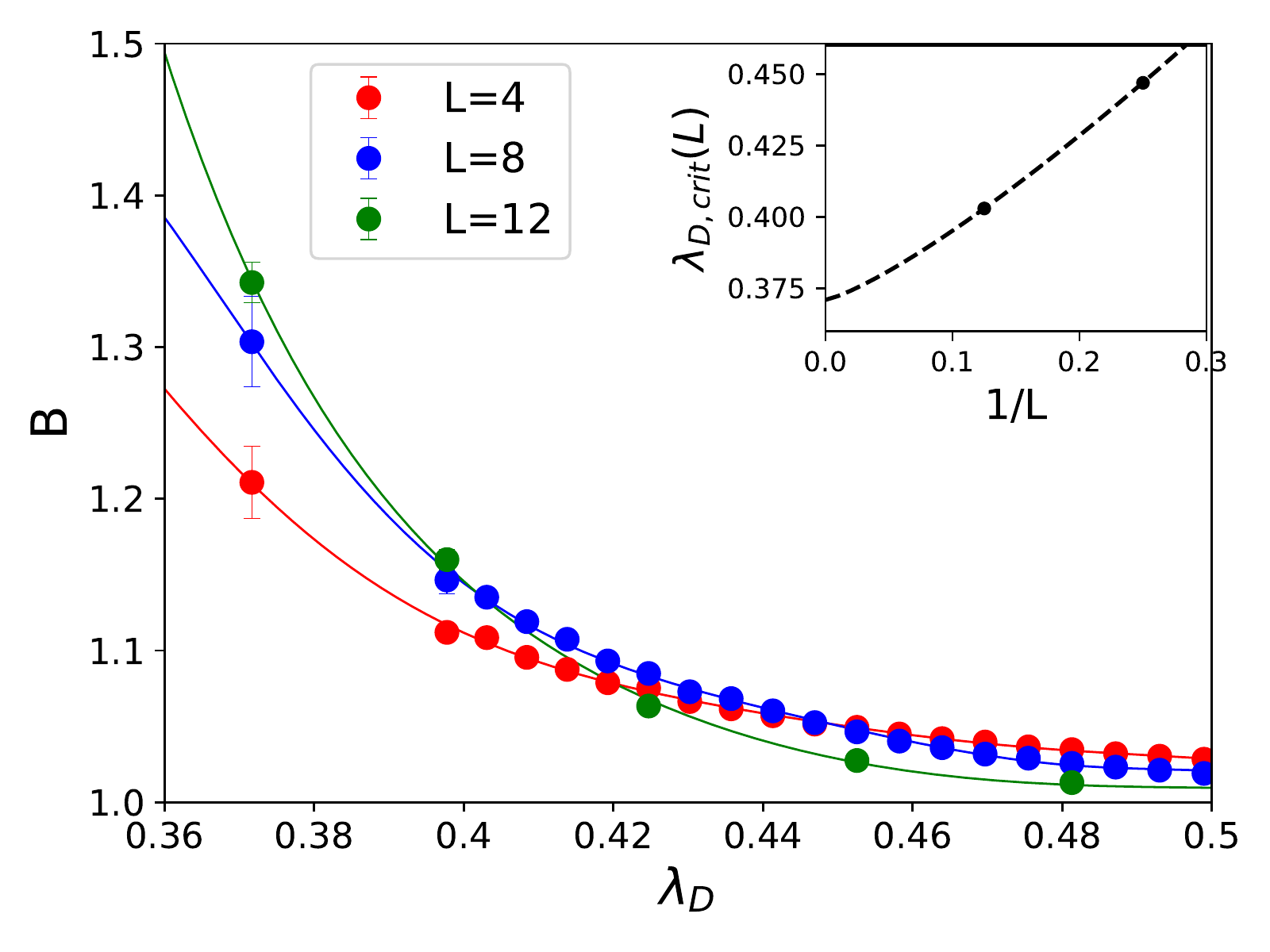}
\caption{
\underbar{Main panel:}
Binder cumulant as a function of EPC strength $\lambda_{D}$ for three
lattice sizes.  Inverse temperature is $\beta =2 \, L$ and $\omega_0$ is fixed at $\omega_0=1.0 \,t$.
\underbar{Inset:}
Extrapolation of the crossings for pairs of sizes
as a function of $1/L$ to get the QCP in the thermodynamic limit.
}
\label{fig:Binder}
\end{figure}

Analysis of the  renormalization group invariant
Binder cumulant\citep{binder81},
\begin{align}
    \mathcal{B} = \frac{3}{2} \left(1 - \frac{1}{3}
\frac{<S_{\rm cdw}^2>}{<S_{\rm cdw}>^2} \right),
\end{align}
can be used to locate the quantum critical point precisely.
Only lattice sizes $L=4n$ where $n$ is an integer
can be used, for other $L$ the Dirac points are not one
of the allowed ${\bf k}$ values and finite size effects are much more
significant.
As exhibited in Fig.~\ref{fig:Binder}, for $L=4,8$ and $12$,
$\mathcal{B}$ exhibits a set of crossings in a range
about $\lambda_{D} \approx 0.4$.
An extrapolation in $1/L$, as shown in the inset of
Fig.~\ref{fig:Binder}, gives $\lambda_{D,{\rm crit}}
=0.371 \pm 0.003$.

\subsection{Phase Diagram}

\begin{figure}[t]
\includegraphics[scale=0.54]{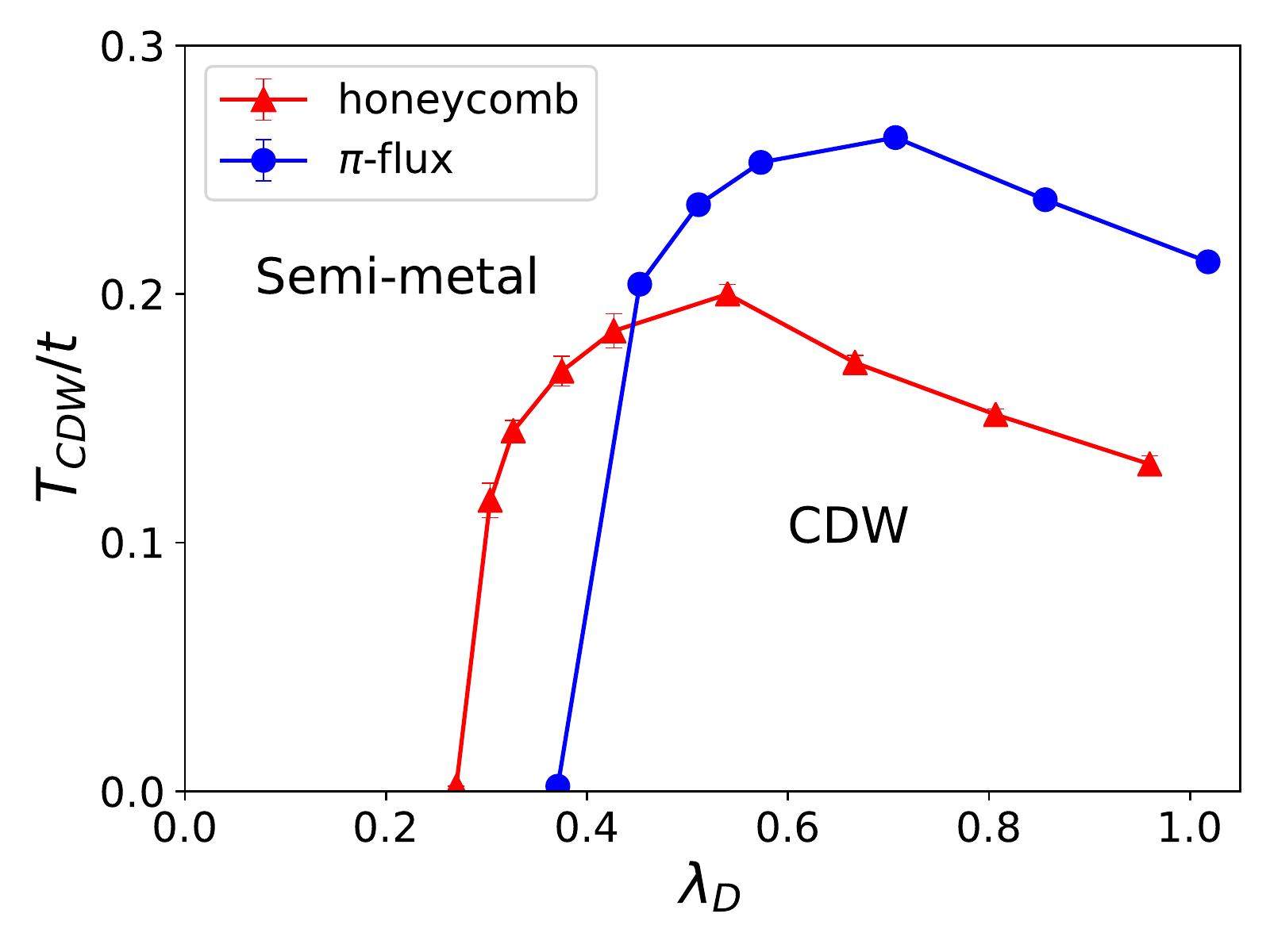}
\caption{Critical temperature $T_{c}$ for CDW phase transition, obtained
from DQMC for both $\pi$-flux phase square lattice (blue line) and the
honeycomb lattice (red line), in a range of coupling strength.
$\lambda_{D}$ is varied by changing $\lambda$ at fixed $\omega_{0}=1.0\,t$
for both models. Quantum critical point is determined using Binder
cumulant analysis (discussed below). Data for the honeycomb lattice are
taken from \citep{zhang19} Error bars are smaller than symbol size for $\pi$-flux data.
}
\label{fig:PD}
\end{figure}

Location of the finite temperature phase boundary, Fig.~\ref{fig:FSS},
and the QCP, Fig.~\ref{fig:Binder},
can be combined into the phase diagram of Fig.~\ref{fig:PD}.
Results for the $\pi$-flux geometry (blue circles) are put in better context
by compared with those of the honeycomb lattice (red triangles).
Data were obtained at fixed $\omega_{0}=1.0\,t$.
In both geometries, phase transitions into CDW order happen only above a finite
$\lambda_{D,{\rm crit}}$.
Beyond
$\lambda_{D,{\rm crit}}$,
$T_{c}$ rises rapidly to its maximal value
before decaying. For $\pi$-flux model, $T_{c}$ reaches
a maximum
$T_{c,{\rm max}}/t \sim 0.26$
at $\lambda_{D} \sim 0.7$, whereas for the honeycomb lattice
$T_{c}$ reaches its maximum
$T_{c,{\rm max}}/t \sim 0.20$
at $\lambda_{D} \sim 0.5$. Similarly
$\lambda_{D,crit}$ for $\pi$-flux is larger than that of the honeycomb
lattice, as $\lambda_{D,crit}=0.42$ and $0.27$ respectively.
When measured in terms of the relative Fermi velocities
$v_{\rm F}=2\,t, \,1.5\,t$ for the $\pi$-flux and honeycomb
respectively,
these values become very similar:
$\lambda_{D,{\rm crit}}/v_{\rm F}=0.21$ and $0.18$
for $\pi$-flux and honeycomb;
$T_{c,{\rm max}}/v_{\rm F}=0.13$ and $0.13$.






\section{VI.  Conclusions}

This paper has determined the quantitative phase diagram for Dirac fermions
interacting with local phonon modes on the $\pi$-flux lattice.
A key feature, shared with the honeycomb geometry, is the presence of
a quantum critical point $\lambda_{D,{\rm crit}}$
below which the system remains a semi-metal down to $T=0$.
The values of $T_c$
and $\lambda_{D,{\rm crit}}$ for the two cases,
when normalized to the Fermi velocities,
agree to within roughly 10\%.

We have also considered the question of whether the properties
of the model can be described in terms of the single ratio
$\lambda^2/\omega_0^2$.  We find that qualitatively this is
indeed the case, but that, quantititively, the charge structure factor
can depend significantly on the individual values of EPC
and phonon frequency, especially in the vicinity of the QCP.
However this more complex behavior is masked
by the fact that $T_c$ rises so rapidly with $\lambda$ in
that region.
In investigating this issue we have
studied substantially smaller values of $\omega_0$
than have typically been investigated in QMC treatments
of the Holstein Hamiltonian.


\vskip0.10in \noindent
\underbar{Acknowledgments:}
The work of Y.-X.Z. and R.T.S. was supported by the grant DE‐SC0014671
funded by the U.S. Department of Energy, Office of Science.  H.G. was supported by NSFC grant
No.~11774019.  The authors would like to thank B. Cohen-Stead and W.-T.
Chiu for useful conversations.

\color{black}

\bibliography{bibpiflux}

\end{document}